\documentclass[11pt]{article} % For LaTeX2e
\usepackage[margin=3.1cm]{geometry}
\usepackage{amsmath,amssymb,amsfonts,graphicx,natbib,color,url}

\title{Dynamic trees for streaming and massive data contexts}

\author{Christoforos Anagnostopoulos\footnote{Corresponding author.}\\
Department of Mathematics, Imperial College London\\
South Kensington Campus, SW7 2AZ.\\
\url{canagnos@imperial.ac.uk}\\
\\
Robert B. Gramacy\\
Booth School of Business, The University of Chicago\\
5807 S.~Woodlawn Ave., Chicago, IL, 60637.\\
\url{rbgramacy@ChicagoBooth.edu}}

\newcommand{\chris}[1]{}
\newcommand{\df}{=_{\text{df}}}
\newcommand{\x}{\mathbf{x}}
\newcommand{\X}{\mathbf{X}}
\newcommand{\y}{\mathbf{y}}

\newcommand{\bm}[1]{\mathbf{#1}}

\newcommand{\bcom}[1]{{#1}}

\newcommand{\mc}[1]{\mathcal{#1}}

\newcommand{\mT}{\mathcal{T}}
\newcommand{\mG}{\mathcal{G}}
\newcommand{\mR}{\mathcal{R}}

\begin{document}

\maketitle
\begin{abstract}
Data collection at a massive scale is becoming ubiquitous in a wide variety of settings, from vast offline databases to streaming real-time information. Learning algorithms deployed in such contexts must rely on single-pass inference, where the data history is never revisited. In streaming contexts, learning must also be temporally adaptive to remain up-to-date against unforeseen changes in the data generating mechanism. Although rapidly growing, the online Bayesian inference literature remains challenged by massive data and transient, evolving data streams. Non-parametric modelling techniques can prove particularly ill-suited, as the complexity of the model is allowed to increase with the sample size. In this work, we take steps to overcome these challenges by porting standard streaming techniques, like data discarding and downweighting, into a fully Bayesian framework via the use of informative priors and active learning heuristics. We showcase our methods by augmenting a modern non-parametric modelling framework, dynamic trees, and illustrate its performance on a number of practical examples. The end product is a powerful streaming regression and classification tool, whose performance compares favourably to the state-of-the-art.

\end{abstract}

\section{Introduction}

In online inference, the objective is to develop a set of update equations that incorporate novel information as it becomes available, without needing to revisit the data history. This results in model fitting algorithms whose space and time complexity remains constant as information accumulates, and can hence operate
% \bcom{, often in real-time,}
in streaming environments featuring continual data arrival, or navigate massive datasets sequentially. Such operational constraints are becoming imperative in certain application areas as \bcom{the scale and real-time nature of }modern data collection continues to grow.

In certain simple cases, online estimation without information loss is possible via exact recursive update formulae, e.g., via conjugate Bayesian updating (see Section \ref{sec:conjugate}). In parametric dynamic modelling, approximate samples from the {filtering} distribution for a variable of interest may be obtained online via sequential Monte Carlo (SMC) techniques, under quite general conditions. In \cite{gramacy2010dtl}, SMC is used in a non-parametric context, where a `particle cloud' of {\em dynamic trees} are employed to track parsimonious regression and classification surfaces as data arrive sequentially. However, the resulting algorithm is not, strictly speaking, {\em online}, since tree moves may require access to the full data history, rather than parametric summaries thereof.  This complication arises as an essential by-product of non-parametric modelling, wherein the complexity of the estimator is allowed to increase with the dataset size.  Therefore, this article recognises that maintaining constant operational cost as new data arrives necessarily requires discarding some (e.g., historical) data.

Specifically, and to help set notation for the remainder of the paper, we consider supervised learning problems with labelled data $(\bm{x}_t,y_t)$, for $t=1,2,\dots,T$, where $T$ is either very large or infinite.  We consider responses $y_t$ which are real-valued (i.e., regression) or categorical (classification).  The $p$-dimensional predictors $\bm{x}_t$ may include real-valued features, as well as binary encodings of categorical ones.  The dynamic tree model, reviewed shortly in Section \ref{sec:dt}, allows sequential non-parametric learning via local adaptation when new data arrive.  However its complexity, and thus computational time/space demands, may grow with the data size $t$.  The only effective way to limit these demands is to sacrifice degrees-of-freedom (DoF) in representation of the fit, and the simplest way to do that is to discard data; that is, to require the trees to work with a subset $w \ll t$ of the data seen so far.

Our primary concern in this paper is managing the information loss entailed in data discarding. First, we propose datapoint \emph{retirement} (Section \ref{sec:retire}), whereby discarded datapoints are partially `remembered' through conjugate informative priors, updated sequentially. This technique is well-suited to trees, which combine non-parametric flexibility with simple parametric models and conjugate priors. Nevertheless, forming new partitions in the tree still requires access to actual datapoints, and consequently data discarding comes at a cost of both information and flexibility.  We show that these costs can be managed, to a surprising extent, by employing the right retirement scheme even when discarding data randomly.  In Section \ref{sec:active_discard}, we further show that borrowing active learning heuristics to prioritise points for retirement, i.e., {\em active discarding}, leads to better performance still.

% The most basic resort in the machine learning community is to assume that information content decays with time, and hence deploy \emph{historical discarding}, aka \emph{sliding windows}, where only the $w$ most recent datapoints, for some fixed window length $w$, are held in memory at any given point in the algorithm. A more sophisticated approach for learning `on a budget' is that of \emph{active learning}, where novel datapoints are disregarded if they are deemed uninformative. In this work, we hybridise the two paradigms into a novel technique we refer to as \emph{active discarding} (Section \ref{sec:active_discard}), where active learning heuristics are deployed to select which datapoints may be removed from memory at the smallest possible cost.

An orthogonal concern 
% to online processing prominent
in streaming data contexts is the need for temporal adaptivity when the concept being learned exhibits {\em drift}.  This is where the data generating mechanism evolves over time in an unknown way. Recursive update formulae will generally require modification to acquire temporally adaptive properties. One common approach is the use of exponential downweighting, whereby the contribution of past datapoints to the algorithm is smoothly downweighted via the use of \emph{forgetting factors}. In Section \ref{sec:forget} we demonstrate how historical data retirement via suitably constructed informative priors can reproduce this effect in the non-parametric dynamic tree modelling context, while remaining fully online.  Using synthetic as well as real datasets, we show how this approach compares favourably against modern alternatives.  The paper concludes with a discussion in Section \ref{sec:conclude}.

% conjugate Bayesian updating, as we describe in Section \ref{sec:forget}, this approach can coincide with the use of suitably constructed informative priors and can hence be deployed without compromising model coherence. We outline how discarded datapoints can be fused into such informative priors, with or without forgetting properties.

\section{Dynamic Trees}\label{sec:dt}
Dynamic trees (DTs) \citep{gramacy2010dtl} are a process--analog of Bayesian treed models \citep{chip:geor:mccu:1998,chip:geor:mccu:2002}.  The model specification is amenable to fast sequential inference by SMC, yielding a predictive surface which organically increases in complexity as more data arrive.  Software is available in the {\tt dynaTree} package \citep{dynaTree} for {\sf R} on CRAN \citep{cranR}, which has been extended to cover the techniques described in this paper. We now review model specification and inference in turn.

%% \subsection{DT Model}\label{sec:model}
\subsection{Bayesian static treed models}
Trees partition the input space $\mathcal{X}$ into hyper-rectangles, referred to as \emph{leaves}, using nested logical rules of the form $(x_i \geq c)$. For instance, the partition $(x_1 \geq 3) \cap (x_2 < -1)$, $(x_1 > 3) \cap (x_2 \geq -1)$ and $(x_1 < 3)$  represent a tree with one internal node, $(x_2 \geq -1)$, and three leaves. We denote by $\eta(\bm{x})$ the unique leaf where $\bm{x}$ belongs to, for any $\bm{x} \in \mathcal{X}$.

Treed models condition the likelihood function on a tree $\mathcal{T}$ and fit one instance of a given simple parametric model per leaf. In this way, a flexible model is built out of simple parametric models $(\theta_\eta)_{\eta \in \mathcal{L}_\mathcal{T}}$, where $\mathcal{L}_\mathcal{T}$ is the set of leaves in $\mathcal{T}$. This flexibility comes at the price of a hard model search and selection problem: that of selecting a suitable tree structure. In the seminal work of \cite{chip:geor:mccu:1998}, a Bayesian solution to this problem was proposed that relied on a generative prior distribution over trees: a leaf node $\eta$ may split with probability $p_{\mathrm{split}}(\mT, \eta) = \alpha(1+D_\eta)^{-\beta}$, where $\alpha, \beta > 0$, and $D_\eta$ is the depth of $\eta$ in the tree $\mT$.
% Partition locations are chosen uniformly over the data input locations $\bm{x}^\eta$ in $\eta$. 
This induces a joint prior via the probability that internal nodes $\mathcal{I}_\mT$ have split and leaves $\mathcal{L}_\mT$ have not: $ \pi(\mT) \propto \prod_{\eta\,\in\,\mc{I}_\mT} p_{\mathrm{split}}(\mT, \eta) \prod_{\eta\,\in\,L_\mT} [1-p_{\mathrm{split}}(\mT, \eta)].
%\label{eq:tprior}
$ %\end{equation}
The specification is completed by employing independent sampling models at the tree leaves: $p(y_1, \dots, y_n | \mathcal{T}, \bm{x}_1,\dots, \bm{x}_n) = \prod_{\eta \in \mathcal{L}_\mT} p(y^\eta | \mathcal{T}, \bm{x}^\eta)$.  Sampling from the posterior proceeds by MCMC, via proposed local changes to $\mT$: so-called {\em grow}, {\em prune}, {\em change}, and {\em swap} ``moves''.  Any data type/model may be used as long as the marginal likelihoods $ p(y^\eta | \mathcal{T}, \bm{x}^\eta)$ are analytic, i.e., as long as their parameters can be integrated out. This is usually facilitated by fully conjugate, scale invariant, default (non-informative) priors, e.g.,:
\begin{equation}
 y \mid \bm{x} \sim N\left(\beta_{\eta(\bm{x})}^T \bm{x} + \mu_{\eta(\bm{x})}, \sigma^2_{\eta(\bm{x})}\right), \; \pi(\beta_{\eta(\bm{x})}, \mu_{\eta(\bm{x})}, \sigma^2_{\eta(\bm{x})}) \propto \frac{1}{\sigma^2_{\eta(\bm{x})}}
\end{equation}
for linear, or, letting $\beta_\eta=\bm{0}$, constant regression leaves. Similarly, multinomial leaves for classification with Dirichlet priors can be employed. These choices yield analytical posteriors \citep{gramacy2010dtl} but also efficient recursive updates for incorporating new datapoints (see Section \ref{sec:conjugate}). 

\subsection{Dynamic Trees} In DTs the ``moves'' are embedded into a process, which describes how old trees mature into new ones when new data arrive. Suppose that $\mT_{t-1}$ represents a set of recursive partitioning rules associated with $\bm{x}^{t-1}$, the set of covariates observed up-to time $t-1$. The fundamental insight underlying the DT process is to view this tree as a \emph{latent state}, evolving according to a state transition probability, $P(\mathcal{T}_t \mid \mathcal{T}_{t-1}, \bm{x}_t)$. The dependence on $\bm{x}_t$ (but not on $y_t$) allows us to consider only moves \emph{local to the current observation}: i.e., pruning or growing can only occur (if at all) for the leaf $\eta(\bm{x}_t)$. This builds computational tractability into the process, as we eitherway need to recompute in that area. Formally, we let: \begin{equation}
 P(\mathcal{T}_t \mid \mathcal{T}_{t-1}, \bm{x}_t)  = 
\begin{cases}
0, &\text{ if }\mathcal{T}_t \text{ is not reachable from }\mathcal{T}_{t-1} \text{ via moves local to }\bm{x}_t\\
p_m \pi(\mathcal{T}_t), &\text{otherwise.}
\end{cases}
\end{equation}
where $p_m$ is the probability of the unique move that can produce $\mathcal{T}_t$ from $\mathcal{T}_{t-1}$, and $\pi$ is the tree prior. We allow three types of moves: grow, prune and stay moves. Each type is considered equiprobable, whereas for grow moves, we choose among all possible split locations by first choosing a dimension $j$ uniformly at random, and splitting $\eta(\bm{x}_t)$ around the location $x_j = \xi$ chosen uniformly at random from the interval formed from the projection of $\eta(\bm{x}_t)$ on the $j$th input dimension. The new observation, $y_t$, completes a stochastic rule for the update $\mT_{t-1} \rightarrow \mT_t$ via $p(y^t | \mT_t, \bm{x}^t)$ for each $\mT_t \in \{\mT_t\}$.  %A s in the static tree model, the leaf % marginal likelihood must be analytic.

  % Maybe mention: An attractive feature of both types of tree model
  % (splitmin and basemax).

%\subsection{Inference}\label{sec:inf}

The DT specification is amenable to Sequential Monte Carlo \citep[e.g.,][]{CarvJohaLopePols2009} inferential mechanics.  At each iteration $t$, the discrete approximation to the tree posterior $\{\mT_{t-1}^{(i)}\}_{i=1}^N$, based on $N$ particles, can be updated to $\{\mT_{t}^{(i)}\}_{i=1}^N$ by {\em resampling} and then {\em propagating}.  Resampling the particles (with replacement) proceeds according to their predictive probability for the next $(\bm{x},y)$ pair, $w_i = p(y_t | \mT_{t-1}^{(i)}, \bm{x}_t)$.  Then, propagating each resampled particle follows the process outlined in 2.2.  Both steps are computationally efficient because they involve only local calculations (requiring only the subtrees of the parent of each $\eta^{(i)}(\bm{x})$). Nevertheless, the particle approximation can shift great distances in posterior space after an update because the data governed by $\eta(\bm{x}_t)^{(i)}$ may differ greatly from one particle to another, and thus so may the weights $w_i$.  This appealing division of labour mimicks the behaviour of an ensemble method without explicitly maintaining one.  As with all particle simulation methods, some Monte Carlo (MC) error will accumulate and, in practice, one must be careful to assess its effect.  Nevertheless, DT out-of-sample performance compares favourably to other nonparametric methods, like Gaussian processes (GPs) regression and classification, but at a fraction of the computational cost \citep{gramacy2010dtl}.

% .  This is true in batch comparisons (i.e., non-sequential data), but especially so in sequential applications like design and optimisation of computer experiments where sequential applications of GP regression or classification models are helpful \cite{gramacy:polson:2011}, but ultimately too slow for industrial strength application.

% A nice byproduct of PL inference for DTs are reliable marginal
% likelihoods via the sequential factorisation $p(y^T |\bm{x}^T) \approx
% \frac{1}{N} \sum_{t=1}^T \log p(y_t | \bm{x}_t, \mT_{t-1}^{(i)})$,
% i.e., using the probabilities calculated in the resample step.
% \cite{gramacy2010dtl} used these to compare leaf model
% specifications (e.g., constant v.~linear) via BFs. 

\section{Datapoint retirement}\label{sec:retire}
At time $t$, the DT algorithm of \cite{gramacy2010dtl} may need to access arbitrary parts of the data history in order to update 
% compute $p(\mT_{t} \mid \mT_{t-1}, \bm{x}^t)$ for each particle
the particles.  Hence, although sequential inference is fast, the method is not technically {\em online}: tree complexity grows as $\log t$, and at every update each of the $\bm{x}^{t} = \left(\bm{x}_1, \dots, \bm{x}_{t}\right)$ locations are candidates for new splitting locations via {\em grow}.  To enable online operation with constant memory requirements, this covariate pool $(\bm{x}^{t})$ must be reduced to a size $w$, constant in $t$. This can only be achieved via data discarding.  Crucially, the analytic/parametric nature of DT leaves enables a large part of any discarded information to be retained in the form of informative leaf priors.  In effect, this yields a \emph{soft} implementation of data discarding, which we refer to as \emph{datapoint retirement}. We show that retirement can preserve the posterior predictive properties of the tree even after data are discarded, and furthermore following subsequent {\em prune} and {\em stay} operations.  The only situation where the loss of data hurts is when new data arrive and demand a more complex tree.  In that case, any retired points would not be available as anchors for new partitions.  Again, since tree operations are local in nature, only the small subtree nearby $\eta(\bm{x}_t)$ is effected by this loss of DoFs, whereas the compliment $\mT_t \setminus \eta(\bm{x}_t)$, i.e., most of the tree, is not affected.

\subsection{Conjugate informative priors at the leaf level}
\label{sec:conjugate}
Consider first a single leaf $\eta\in \mT_t$ in which we have already retired some data.  That is, suppose we have discarded $(\bm{x}_s,y_s)_{\{s\}}$ which was in $\eta$ in $\mT_{t'}$ at some time $t' \leq t$. The information in this data can be `remembered' by taking the leaf-specific prior, $\pi(\theta_\eta)$, to be the posterior of $\theta_\eta$ given (only) the retired data.  Suppressing the $\eta$ subscript, we may take $ %\[
\pi(\theta) \df P\left( \theta \mid (\bm{x}_{s},y_{s})_{\{s\}}\right) \propto L\left(\theta; (\bm{x}_{s},y_{s})_{\{s\}}\right)\pi_0(\theta) $ % \]
where $\pi_0(\theta)$ is a baseline non-informative prior employed at all leaves. % in the absence of retired data.  
The {\em active data} in $\eta$, i.e., the points which have not been retired, enter into the likelihood in the usual way to form the leaf posterior.  % We shall return to this posterior, and corresponding posterior predictive distribution momentarily.

It is fine to {\em define} retirement in this way, but more important to argue that such retired information can be updated loslessly, and in a computationally efficient way.  Suppose we wish to retire one more datapoint, $(\bm{x}_r,y_r)$.  
Consider the following recursive updating equation:
\begin{align}
\label{eq:tractable}\pi^{(\text{new})}(\theta)  &\df P\left(\theta \mid (\bm{x}_s,y_s)_{\{s\}, \bcom{r}}\right)\propto L(\theta; \bm{x}_r,y_r) P\left(\theta \mid (\bm{x}_{s},y_s)_{\{s\}}\right) .
\end{align}
As shown below, the calculation in (\ref{eq:tractable}) is tractable whenever conjugate priors are employed.

Consider first the linear regression model, $\mathbf{y} \sim N(\mathbf{X}\beta, \sigma^2 \mathbf{I})$, where $\y = (y_s)_{\{s\}}$ is the retired response data, and $\X$ the retired \emph{augmented} design matrix, i.e., whose rows are like $[1,\x_s']'$, so that $\beta_1$ represents an intercept. With $\pi_0(\beta,\sigma^2) \propto \frac{1}{\sigma^2}$, we obtain: \[ \pi(\beta,\sigma^2) \df P(\beta,\sigma^2 \mid \y, \mathbf{X})= \text{NIG}(\nu/2, s\nu/2,\beta,\mG^{-1}) \] where NIG stands for Normal-Inverse-Gamma, and assuming the Grahm matrix $\mG =\X'\X$ is invertible and denoting $Xy = \X'\y, \; r = \y'\y$, we have $\nu = n-p, \; \beta = \mG^{-1} Xy$, and $s^2 = \frac{1}{\nu}(r - \mR)$, where $ \mR = \beta\ \mG^{-1} \beta$.  Having discarded $(y_s,\bm{x}_s)_{\{s\}}$, we can still afford to keep in memory the values of the above statistics, as, crucially, their dimension does not grow with $|\{s\}|$. Updating the prior to incorporate an additional retiree $(y_r,\bm{x}_r)$ is easy: \begin{align} \label{eq:linear_nonconvex} \mG^{\mathrm{(new)}} &= \mG + X_r' X_r, & Xy^{(\mathrm{new})} &= Xy + \X_r'\y_r, &s^{\text{(new)}} &= s + y_r'y_r, & \nu^{\text{(new)}} = \nu + 1.  \end{align}

% Having presented in detail the linear model, we now present in brief the constant and multinomial cases, % referring to \cite{kendall2004ats} for more details. 
The constant leaf model may be obtained as a special case of the above, where $\mathbf{x}^{\star} = 1$, $\mG = \nu$ and $\beta = \mu$. %, i.e., the intercept now represents the mean of $y$.
For the multinomial model, the discarded response values $y_s$ may be represented as indicator vectors $\mathbf{z}_s$, where $z_{js} = \mathbf{1}(y_s = j)$. The natural conjugate here is the \emph{Dirichlet} $D(\mathbf{a})$. The hyperparameter vector $\bm{a}$ may be interpreted as counts, and is updated in the obvious manner, namely $% \[
\bm{a}^{(new)} = \bm{a} + z_r $ % \]
where $z_{jm} = \mathbf{1}(y_r = j)$. A sensible baseline is $\bm{a}_0 = (1,1,\dots,1)$.  See \cite{kendall2004ats} for more details.
% , and also find it useful to employ a normalised representation of $\bm{a}$, given by $\bm{t} = \bm{a}/\sum_j a_{j}$.

Unfolding the updating equations (\ref{eq:tractable}) and (\ref{eq:linear_nonconvex}) makes it apparent that retirement preserves the posterior distribution.  Specifically, the posterior probability of parameters $\theta$, given the active (non-retired) data still in $\eta$ is
\[
\pi(\theta | \bm{x}^\eta, y^\eta) \propto L(\theta; x^\eta, y^\eta) \pi(\theta) 
\propto L(\theta; x^\eta, y^\eta)   L(\theta; (\bm{x}_{s},y_{s})_{\{s\}})\pi_0(\theta) 
= L(\theta; x^{\eta'}, y^{\eta'}) \pi_0(\theta),
\]
where $\eta'$ is $\eta$ without having retired $ (\bm{x}_{s},y_{s})_{\{s\}}$.  Since the posteriors are unchanged, so are the posterior predictive distributions and the marginal likelihoods required for the SMC updates.  Note that new data $(\bm{x}_{t+1}, y_{t+1})$ which do not update a particular node $\eta \in \mT_t \rightarrow \mT_{t+1}$ do not change the properties of the posterior local to the region of the input space demarcated by $\eta$.  It is as if the retired data were never discarded.  Only where updates demand modifications of the tree local to $\eta$ is the loss in DoF felt.  We argue in Section \ref{sec:tree_retire} that this impact can be limited to operations which {\em grow} the tree locally.  Cleverly choosing which points to retire can further mitigate the impact of discarding (see Section \ref{sec:active_discard}).

\subsection{Managing informative priors at the tree level}\label{sec:tree_retire}
Intuitively, DTs with retirement manage two types of information: a non-parametric memory comprising an {active data pool} of constant size $w \ll t$, which forms the leaf likelihoods; and a parametric memory consisting of possibly informative leaf priors. The algorithm we propose proceeds as follows.  At time $t$, add the $t^{\mathrm{th}}$ datapoint to the active pool, and update the model by SMC exactly as explained in Section \ref{sec:dt}. Then, if $t$ exceeds $w$, also select some datapoint, $(\bm{x}_r,y_r)$, and discard it from the active pool (see Section \ref{sec:active_discard} for selection criteria), having first updated the associated leaf prior for $\eta(\bm{x}_r)^{(i)}$, for each particle $i=1,\dots, N$, to `remember' the information present in $(\bm{x}_r,y_r)$.  This shifts information from the likelihood part of the posterior to the prior, exactly preserving the time-$t$ posterior predictive distribution and marginal likelihood for every leaf in every tree.\footnote{In fact, every data point under active management can [in a certain limited sense] be retired without information loss.}

The situation changes when the next data point $(\bm{x}_{t+1}, y_{y+1})$ arrives.  Recall that the DT update chooses between {\em stay}, {\em prune}, or {\em grow} nearby each $\eta(\bm{x}_{t+1})^{(i)}$.
% Stay moves clearly do not cause a problem, but 
Grow and prune moves are affected by the absence of the retired data from the active data pool. In particular, the tree cannot grow if there are no active data candidates to split upon. This informs our assessment of retiree selection criteria in Section 4, as it makes sense not to discard points in parts of the input space where we expect the tree to require further DoFs. Moreover, we recognise that the stochastic choice between the three DT moves depends both upon the likelihood, and retired (prior) information local to $\eta(\bm{x}_{t+1})^{(i)}$, so that the way that prior information propagates after a prune, or grow move, matters. The original DT model dictates how likelihood information (i.e., resulting from active data) propagates for each move.  We must provide a commensurate propagation for the retired information to ensure that the resulting online trees stay close to their full data counterparts.

If  a {\em stay} move is chosen stochastically, no further action is required: retiring data has no effect on the posterior.  When nodes are grown or pruned, the retiring mechanism itself, which dictates how informative priors can salvage discarded likelihood information, suggests a method for splitting and combining that information.
% and thus for handling priors built on retired data.
 Following a {\em prune}, retired information from the pruned leaves, $\eta$ and its sibling $S(\eta)$, must be pooled into the new leaf prior positioned at the parent $P(\eta)$.  
Conjugate updating suggests the following additive rule:
\begin{align*}
\mG^{P(\eta)} &= \mG^{\eta} + \mG^{S(\eta)}, & Xy^{P(\eta)} &= Xy^{\eta} + Xy^{S(\eta)} 
& s^{P(\eta)} &= s^{\eta} + s^{S(\eta)}, &\nu^{P(\eta)} &= \nu^{\eta} + \nu^{S(\eta)}.  
\end{align*}
Note that this does not require access to the actual retired datapoints, and would result in the identical posterior even if the data had not been discarded.

A sensible {\em grow} move can be derived by reversing this logic.  We suggest letting both novel child leaves $\ell(\eta)$ and $r(\eta)$ inherit the parent prior, but split its strength $\nu^{\eta}$ between them at proportions equal to the active data proportions in each child.   Let $\alpha =  \frac{|\ell(\eta)|}{|\eta|}$. Then,
\begin{align*}
\nu_{\ell(\eta)} &= \alpha \nu_\eta, &\mG^{\ell(\eta)} &= \alpha \mG^{\eta}, & 
Xy^{\ell(\eta)} &= \alpha Xy^{\eta}, & s^{\ell(\eta)} & = \alpha s^{\eta}, \\
\nu_{r(\eta)} &= (1-\alpha) \nu_\eta, &\mG^{r(\eta)} &= (1-\alpha) \mG^{\eta}, & 
Xy^{r(\eta)} &= (1-\alpha) Xy^{\eta}, & s^{r(\eta)} & = (1-\alpha) s^{\eta}.
\end{align*}
In other words, the new child priors share the retired information of the parents with weight proportional to the number of active data points they manage relative to the parent.  This preserves the total strength of retired information, preserves the balance between active data and parametric memory, and is {\em reversible}: subsequent {\em prune} operations will exactly undo the partitioned prior, combining it into the same prior sufficient statistics at the parent.

This brings to light a second cost to discarding data, the first being a loss of candidates for future partitioning. Nodes grown using priors built from retired points lack specific location information from the actual retired $(\bm{x}_s,y_s)$ pairs.  Therefore newly grown leaves must necessarily compromise between explaining the new data, e.g., $(\bm{x}_{t+1}, y_{t+1})$, with immediately local data active data to $\eta(\bm{x})_{t+1}$, and information from retired points with less localised influence.  The weight of each component in the compromise is $|\eta|/(|\eta| + \nu_\eta)$ and $\nu_\eta/(|\eta| + \nu_\eta)$, respectively.  Eventually as $t$ grows, with $w \ll t$ staying constant, retired information naturally dominates, precluding new grows even when active partitioning candidates exist.  This means that while the hierarchical way in which retired data filters through to inference (and prediction) at the leaves is sensible, it is doubly-important that data points in parts of the input space where the response is very complex should not be discarded. 

\section{Active discarding}\label{sec:active_discard}
% In Section \ref{sec:retire}, we explained how retirement softens the blow of data discarding by moving leaf sufficient statistics into the leaf priors.  Indeed, it does so in an optimal way in parts of the tree space which are inert to future observations, since the posterior predictive distribution is left unchanged in leaves that remain untouched by future data.  We also observed that in leaves that are touched by future data, the degrees of freedom lost to retired points % are missed, particularly for {\em grow} which relies on the active data pool for possible splitting locations. Consequently,

It matters which data points are chosen for retirement, so it is desirable 
% Therefore, we want 
to retire datapoints that will be of ``less'' use to the model going forward. In the case of a drifting concepts, retiring \emph{historically}, i.e., retiring the oldest datapoints, may be sensible.  We address this in Section \ref{sec:forget}.  Here we consider static concepts, or in other words i.i.d.~data.  We formulate the choice of which active data points to retire as an {\em active discarding} (AD) problem by borrowing (and reversing) techniques from the {\em active learning} (AL) literature.  Regression and classification models separately, as they require different AD techniques.  We shall argue that in both cases AD is, in fact, easier than AL since DTs enable thrifty analytic calculations not previously possible, which are easily updated within the SMC.

\subsection{Active discarding for regression}

{\em Active learning} (AL) procedures are sequential decision heuristics for choosing data to add to the design, usually with the aim of minimising prediction error.  Two common AL heuristics are active learning MacKay \citep[ALM]{mackay:1992} and active learning Cohn \citep[ALC]{cohn:1996}.  They were popularised in the modern nonparametric regression literature \citep{seo:00} using GPs, and subsequently ported to DTs \citep{gramacy2010dtl}.  An ALM scheme selects new inputs $\bm{x}^{\star}$ with maximum variance for $y(\bm{x}^{\star})$, whereas ALC chooses $\bm{x}^{\star}$ to maximise the expected reduction in predictive variance averaged over the input space. Both approximate maximum expected information designs in certain cases.  ALC is computationally more demanding than ALM, requiring an integral over a set of reference locations that can be expensive to approximate numerically for most models.  But it leads to better exploration when used with nonstationary models like DTs because it concentrates sampled points near to where the response surface is changing most rapidly \citep{gramacy2010dtl}.  ALM has the disadvantage that it does not cope well with heteroskedastic data (i.e., input-dependent noise).  It can end up favouring regions of high noise rather than high model uncertainty.  Both are sensitive to the choice of (and density of) a search grid over which the variance statistics are evaluated.

Our first simplification when porting AL to AD is to recognise that no grids are needed.  We focus on the ALC statistic here because it is generally preferred, but also to illustrate how the integrals required are actually very tractable with DTs, which is not true in general. The AD program is to evaluate the ALC statistic at each active data location, and choose the smallest one for discarding.  AL, by contrast, prefers large ALC statistics to augment the design.  We focus on the linear leaf model, as the constant model may be derived as a special case.
%
%First ALC under
%the constant leaf model.  
%\begin{equation}
%\Delta \sigma^2_x(\bm{z} | \mT) = \Delta
%\sigma^2_x(\bm{z}|\eta) \equiv
%\sigma^2(\bm{z} | \eta) - \sigma^2_x(\bm{z}|\eta) =
%\frac{s_{\eta}^2}{|\eta|-3} \times
%\frac{\left(\frac{1}{|\eta|}\right)^2}{1 + \frac{1}{|\eta|}}  \label{eq:alcc}
%\end{equation}
% Since %(\ref{eq:alcc} is constant (with respect to $\bm{x}$), 
%integrating over the reference locations $\bm{z}$, to get the expected
%reduction in variance over the entire input space, is easy:
%\[
%\Delta \sigma^2(\bm{x}) = \int_{\mathbb{R}^d} \Delta
%\sigma^2_{\bm{x}}(\bm{z})\,d \bm{z} = A_\eta \times
%\frac{s_{\eta}^2}{|\eta|-3} \times
%\frac{\left(\frac{1}{|\eta|}\right)^2}{1 + \frac{1}{|\eta|}},
%\]
%where $A_\eta$ is the area of the rectangle demarcated by $\eta$.  
%
% This expression is true whether learning or discarding, however for AL we must evaluate $\Delta \sigma^2(x)$ over a dense candidate grid of $x$'s.  For AD we need only consider the current active data locations.
%
% The linear leaf model requires a bit more work.
For an active data location $\bm{x}$ and (any) reference location $\bm{z}$,
the reduction in variance at $\bm{z}$
given that $\bm{x}$ is in the design, and a tree $\mT$ is given by (see \cite{gramacy2010dtl}):
\[
\Delta
\sigma^2_{\bm{x}}(\bm{z}|\mT) = \Delta
\sigma^2_{\bm{x}}(\bm{z}|\eta) \equiv
\sigma^2(\bm{z} | \eta) - \sigma^2_x(\bm{z}|\eta) =
\frac{s_{\eta}^2 - \mathcal{R}_{\eta}}{|\eta|-m-3} \times
\frac{\left(\frac{1}{|\eta|} +\bm{z}^\prime 
\mathcal{G}^{-1}_{\eta}\bm{x}\right)^2}{
1 + \frac{1}{|\eta|} + \bm{x}^\prime 
\mathcal{G}^{-1}_{\eta}\bm{x}},
\]
when {\em both} $\bm{x}$ and $\bm{z}$ are in $\eta\in \mathcal{L}_\mathcal{T}$, and zero otherwise.  
% The $s_\eta^2$ term is the leaf sum of squares, $|\eta|$ is the number of data points in the leaf, and $\mathcal{G}_\eta$ is the Grahm matrix.  In each case these quantities must be derived from the for the active {\em and} retired data in $\eta$.
This expression is valid whether learning or discarding, however AL requires evaluating $\Delta \sigma^2(x)$ over a dense candidate grid of $x$'s.  AD need only consider the current active data locations, which can represent a dramatic savings in computational cost.

Integrating over $\bm{z}$ gives:
\[
\Delta \sigma^2(\bm{x}) = \int_{\mathbb{R}^d} \Delta
\sigma^2_{\bm{x}}(\bm{z})\,d\bm{z} = 
\frac{s_{\eta}^2 - \mathcal{R}_{\eta}}{(|\eta|-m-3)(
1 + \frac{1}{|\eta|} + \bm{x}^\prime 
\mathcal{G}^{-1}_{\eta}\bm{x})} 
\times \int_{\eta}
\left(\frac{1}{|\eta|} +\bm{z}^\prime 
\mathcal{G}^{-1}_{\eta} \bm{x}\right)^2 \, d\bm{y}.
\]
The integral that remains, over the rectangular region $\eta$,
is tedious to write out but has a trivial $O(m^2)$ implementation. 
Let the $m$-rectangle $\eta$ be described by $\{(a_i,b_i)\}^m$.  Then,
\begin{align*}
  \int_{a_1}^{b_1} & \cdots \int_{a_m}^{b_m} \left(c + \sum_{i=1}^m \tilde{z}_i x_i \right)^2 dz_1 \cdots dz_m = A_\eta c^2
 + c \sum_i \left( \prod_{k\neq i}  (b_k-a_k) \right) x_i (b_i^2 -
a_i^2) \\
 &\quad + \sum_i \left(\prod_{k\neq i}
  (b_k-a_k)\right)\frac{x_i^2}{3}(b_i^3-a_i^3) +  \sum_i \sum_{j < i} \left(\prod_{k\neq i,j} (b_k-a_k)\right) 
\frac{x_i x_j}{2} (b_i^2- a_i^2)(b_j^2 - a_j^2), 
\end{align*}
where $\tilde{\bm{z}} = \bm{z}' \mathcal{G}^{-1}_\eta$, and $c = 1/|\eta|$.  A general-purpose numerical version via sums using $R$ reference locations $\bm{z}$---previously  the state of the art \citep{gramacy2010dtl}---requires $O(Rm)$ computation with $R$ growing exponentially in $m$ for reasonable accuracy.  Observe that the rectangular leaf regions generated by the trees is key.  In the case of other partition models (like Voronoi tessellation models), this analytical integration would not be possible.

In repeated applications of ALC for AD, we observe that the active points
that remain tend shuffle themselves so that they cluster near the
high posterior partitioning boundaries, which makes sense because these are the locations where the predictive surface is changing the fastest.  The number of
such locations depends on the number of active data points
allowed, $w$.
\begin{figure}[ht!]
\centering
\includegraphics[scale=0.45,trim=30 10 30 10]{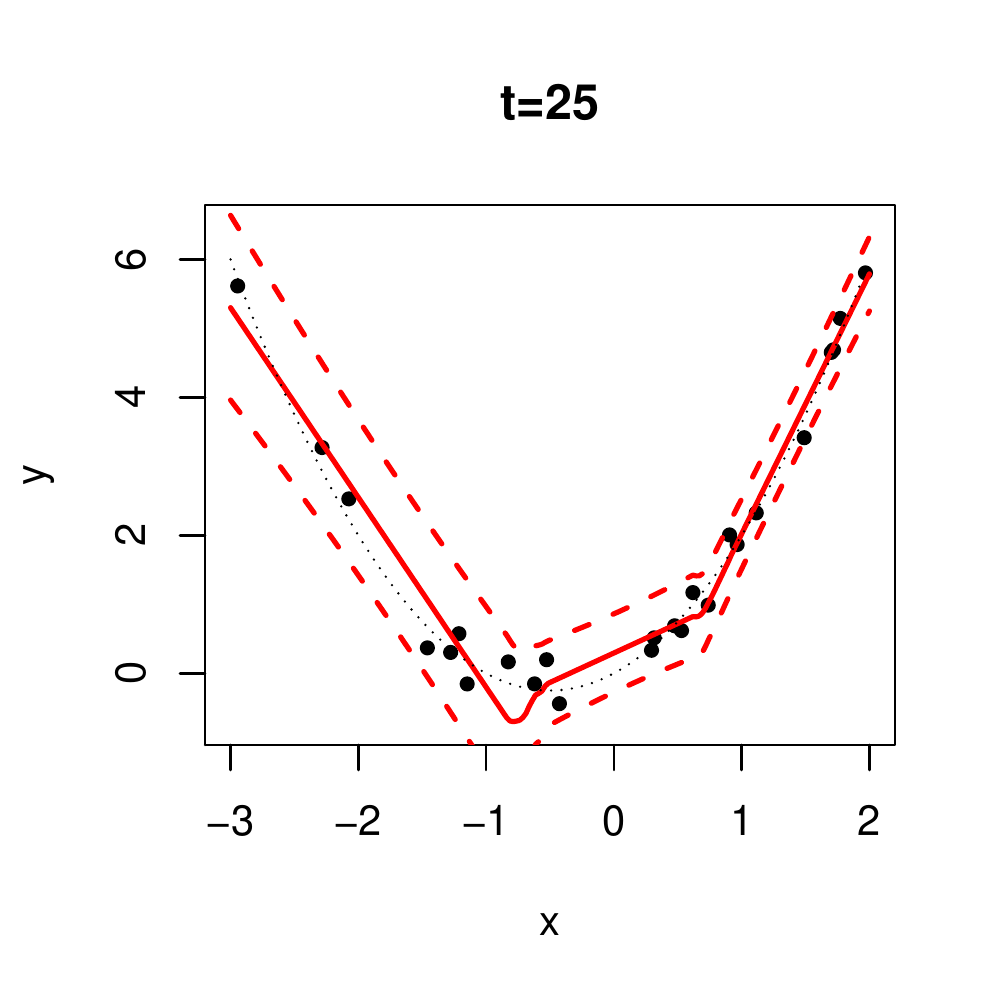}
\includegraphics[scale=0.45,trim=53 10 30 10,clip=TRUE]{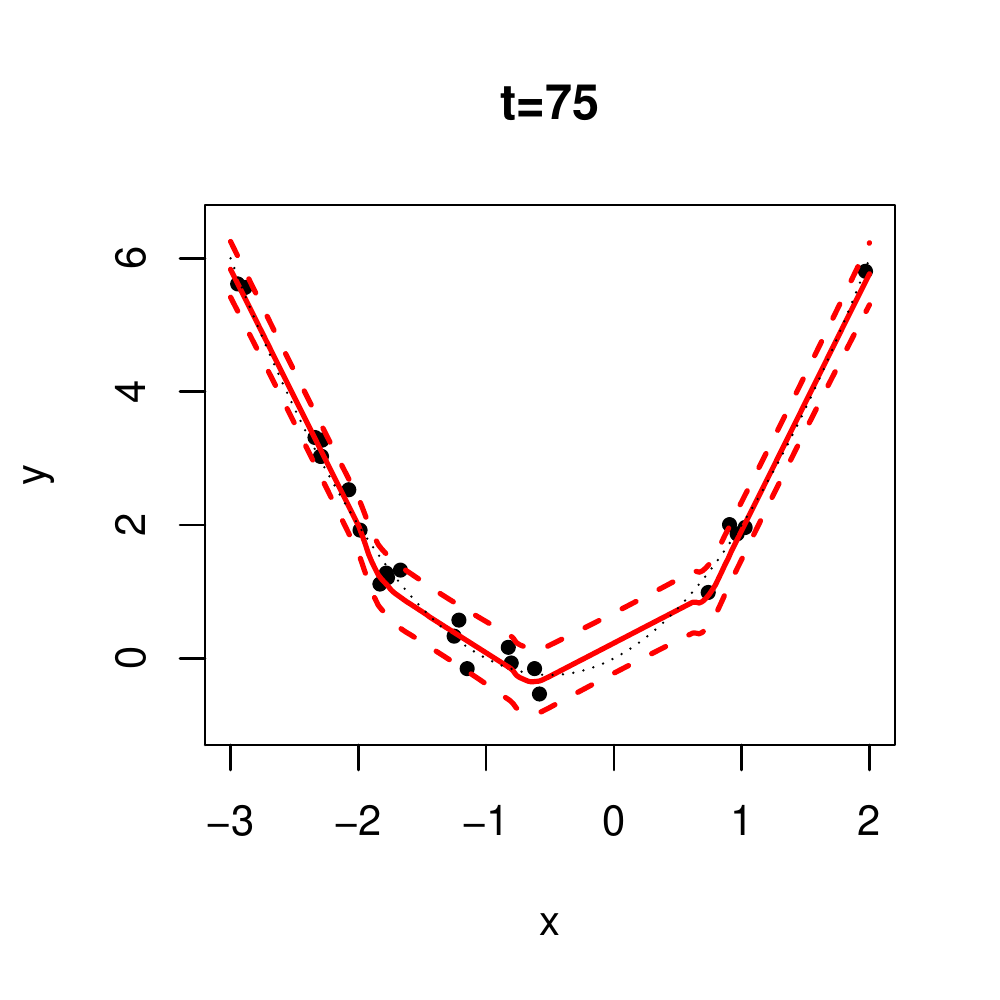}
\includegraphics[scale=0.45,trim=53 10 30 10,clip=TRUE]{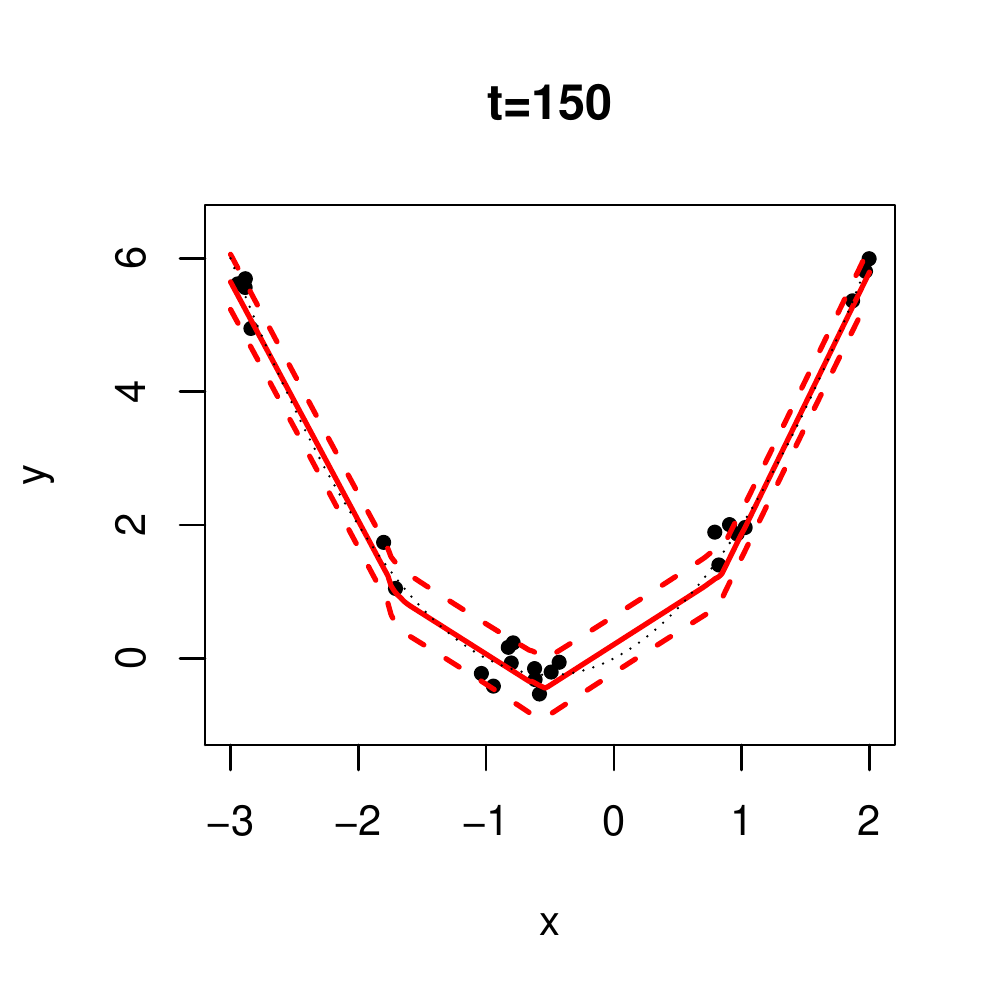}
\includegraphics[scale=0.45,trim=53 10 30 10,clip=TRUE]{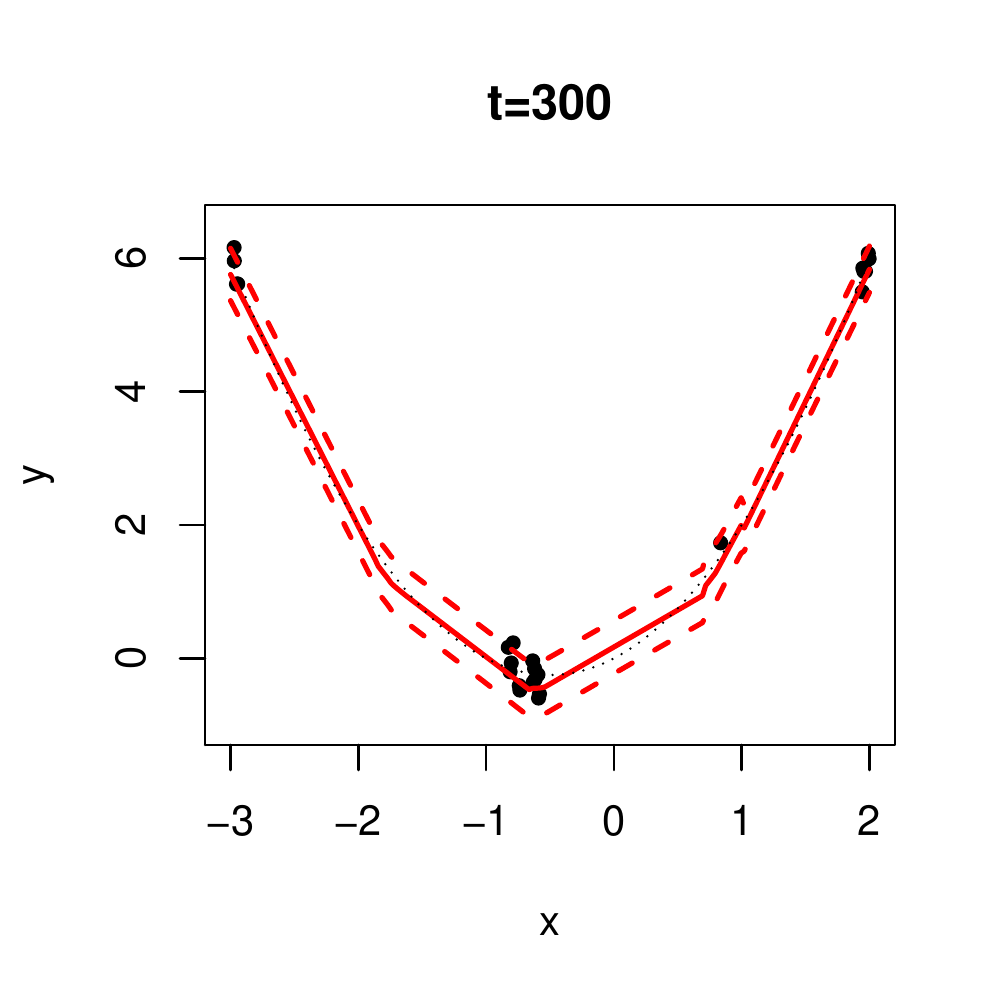}
\caption{Snapshots of active data (25 points) and predictive surfaces
  spanning 275 retirement/updating rounds;  $Y(x) = x + x^2 +
  \varepsilon$, $\varepsilon \sim \mathcal{N}(0,1)$.}
\label{f:alcparab}
\end{figure}
As an illustration, consider the simple example where the response is a parabolic function, which must be learned sequentially via $x$-data sampled uniformly in $(-3,2)$, with $w=25$.  The initial 25, before any retiring, are shown in the first panel.  Each updating round then proceeds with one retirement followed by one new pair, and subsequent SMC update.  %Notice how the active points shift towards likely partition boundaries. 
Since the implementation requires at least five points in each leaf, seeing four regimes emerge is perhaps not surprising.  By $t=150$, the third pane, the ability to learn about the mean with just 25 degrees of freedom is saturated, but it is possible to improve on the variance (shown as errorbars), which are indeed smaller in the final $t=300$ pane.  Eventually, the points will cluster at the ends because that is where the response is changing most rapidly, and indeed the derivative is highest there (in absolute value).

\subsection{Active discarding for classification}

For classification, predictive entropy is an obvious AL heuristic.  Given a predictive surface comprised of probabilities $p_\ell(\bm{x})$ for each class $\ell$, from DTs or otherwise, the predictive entropy at $\bm{x}$ is $-\sum_{\ell} p_\ell(\bm{x}) \log p_\ell(\bm{x})$. Entropy can be an optimal method for measuring predictive uncertainty, but that does not mean it is good for AL.  Many authors \citep[e.g.,][]{joshi:2009} have observed that it can be too greedy: entropy can be very high near the best explored class boundaries.  % In other words, the higher the certainty about a boundary, the higher the entropy there, making it poor for exploring the full boundary set.
Several, largely unsatisfactory, remedies have been suggested in the literature.  Fortunately, no remedy is required for the AD analog, which focuses on the lowest entropy active data, a finite set.  The discarded points will tend to be far into the class interior, where they can be safely subsumed into the prior. 
%, whereas the retained (non-retired) points will be no closer to the boundary than they already are: the arrival of novel datapoints is still governed by the sampling distribution over $\bm{x}_t$.  
Their spacing and shifting of the active pool is quite similar to discarding by ALC for regression, and so we do not illustrate it here.  % Empirical results are deferred to Section \ref{sec:adresults}.

\subsection{Fast local updates of active discarding statistics with
  trees}

%\bcom{At the level of the full particle cloud, we average the AD statistics of each individual particle over all particles, for each active datapoint, and at each timestep remove the one with the smallest such statistic.} 
The divide and conquer nature of trees---whose posterior distribution is approximated by thrifty, local, particle updates---allows AD statistics to be updated cheaply too.  If each leaf node stores its own AD statistics, %then rather than re-calculating ALC or entropy for every active input, after each update step,
it suffices to update only the ones in leaf nodes which have been modified, as described below.  Any recalculated statistics can then be subsumed into a global, particle averaged, version. % (after first subtracting off the old values).
Note that no updates to the AD statistics are needed when a point is retired since the predictive distributions are unchanged.

When a new datapoint $(\bm{x},y)$ arrives, the posterior undergoes two types of changes: resample then propagate.  In the resample step the discrete particle distribution changes, although the trees therein do not change.  Therefore, each discarded particle must have its AD statistics (stored at the leaves) subtracted from the full particle tally.  Then each correspondingly duplicated particle can have its AD statistics added in.  No new integrals
(for ALC) or entropy calculations (for classification) are needed.  In the propagate step, each particle undergoes a change local to $\eta(\bm{x})^{(i)} \in \mT^{(i)}_t$.  This requires first calculating the AD statistic for the new $(\bm{x},y)$ for each $\eta^{(i)}(\bm{x})$, before the dynamic update occurs, and then swapping it into the particle average.  
New integrations, etc., need evaluating here.   Then, each non-{\em stay} dynamic update triggers swap of the old AD statistics in $\eta^{(i)}(\bm{x})$ for freshly re-calculated ones 
from the leaf node(s) in $\mT_{t+1}^{(i)}$.  The total computational cost is in $O(m^2N)$ for incorporating $(\bm{x}, y)$ into $N$ particles, plus $O(m^2\sum_{i=1}^N |\eta^{(i)}(x)|)$ to update the leaves.\footnote{One might imagine a
  thriftier, but harder to implement version, which waits until the
  end to calculate the AD statistic for the new point
  $(x,y)$.  But it would have the same computational order.}

\subsection{Empirical results}
\label{sec:adresults}

Here we explore the benefit of AD over simpler heuristics, like random discarding and subsetted data estimators, by making predictive comparisons on benchmark regression and classification data.  To focus the discussion on our key objective for this section, we employ moderate data sample sizes in order to allow a comparison to full-data versions of DTs, and assess the impact of data discarding on performance. \bcom{In particular, we do not repeat here a comparison of full-data DTs to competitors, which may be found in \cite{gramacy2010dtl}, but emphasise that discarding enables DTs to operate on (arbitrarily long) data streams, where the original DTs, as well as their main GP-based competitors, will eventually become intractable. This is better illustrated by the use of massive and streaming classification datasets in Section \ref{sec:forget}.}

\subsubsection*{Simple synthetic regression data}

We first consider data originally used to illustrate multivariate adaptive regression splines (MARS) \citep{fried:1991}, and then to demonstrate the competitiveness of DTs relative to modern (batch) nonparametric models \citep{gramacy2010dtl}. The response is $10 \sin(\pi x_1 x_2) + 20(x_3 - 0.5)^2 + 10x_4 + 5 x_5$ plus $\mathcal{N}(0,1)$ additive error. Inputs $\bm{x}$ are random in $[0,1]^5$.  We considered four estimators: one based on 200 pairs (ORIG), one based on 1800 more for 2000 total (FULL), and two online versions using either random (ORAND) or ALC (OALC) retiring to keep the total active data set limited to $w=200$.   ORIG is intended as a lower benchmark, representing a na\"ive fixed-budget method;  FULL
is at the upper end.
%\ bcom{ORIG is a lower benchmark, as it represents the most na\:ive way of maintaining a fixed budget of active datapoints. ORAND is just as na\:ive when it comes to its selection criterion, but gets to see the full dataset, which can lead to more mature trees, especially in the presence of retirement where information builds up in the leaf priors. Finally, FULL represents an upper benchmark in this context, as it is allowed to employ all the information in the data.} 
The full experiment was comprised of 100 repeats in a MC fashion, each with new random training sets, and random testing sets of size 1000.  $N=1000$ particles and a linear leaf model were used throughout.  Similar results were obtained for the constant model.

\begin{figure}[ht!]
%\vspace{0.25cm}
\begin{minipage}{8.5cm}
\includegraphics[scale=0.32,trim=20 50 0 50]{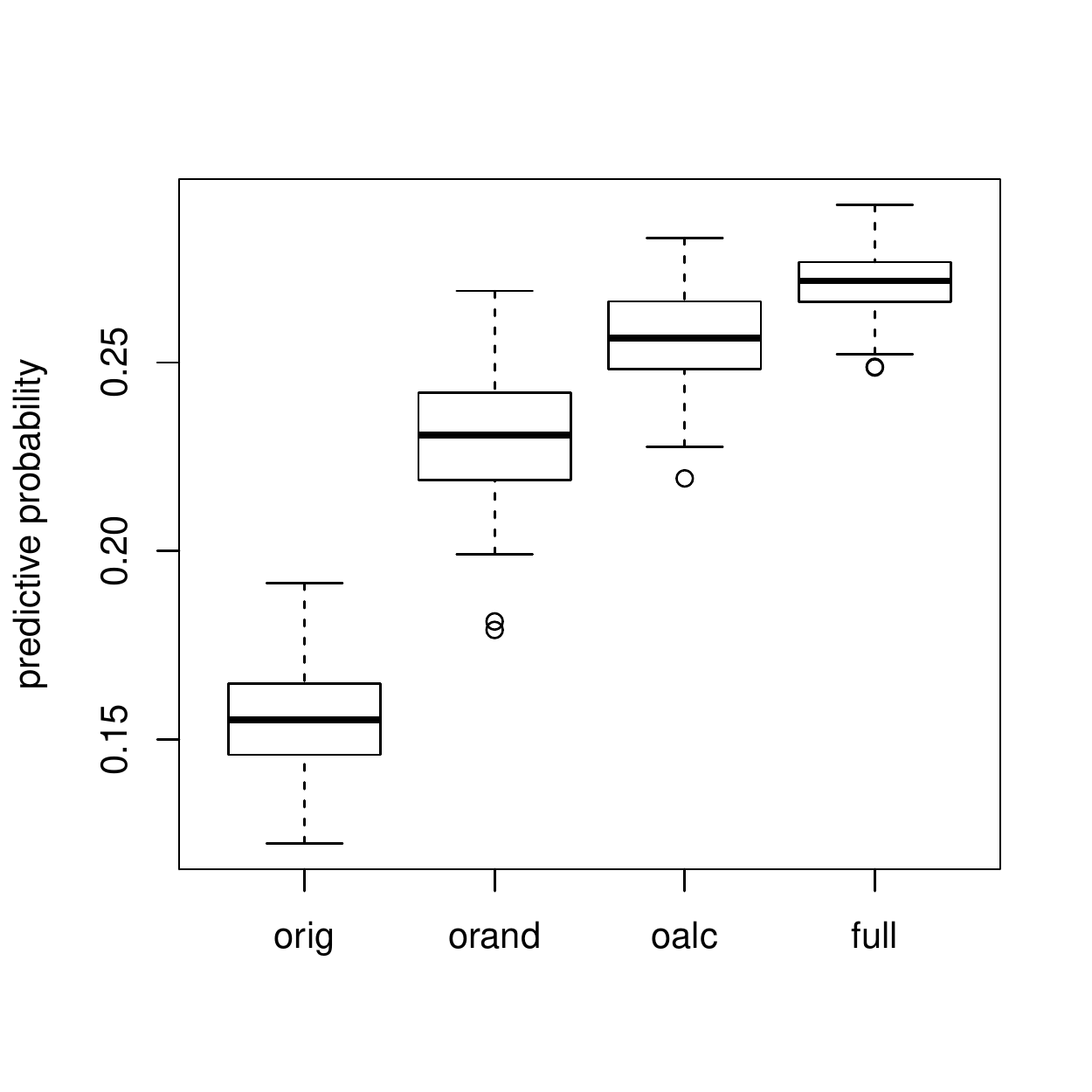}
\includegraphics[scale=0.32,trim=20 50 0 50]{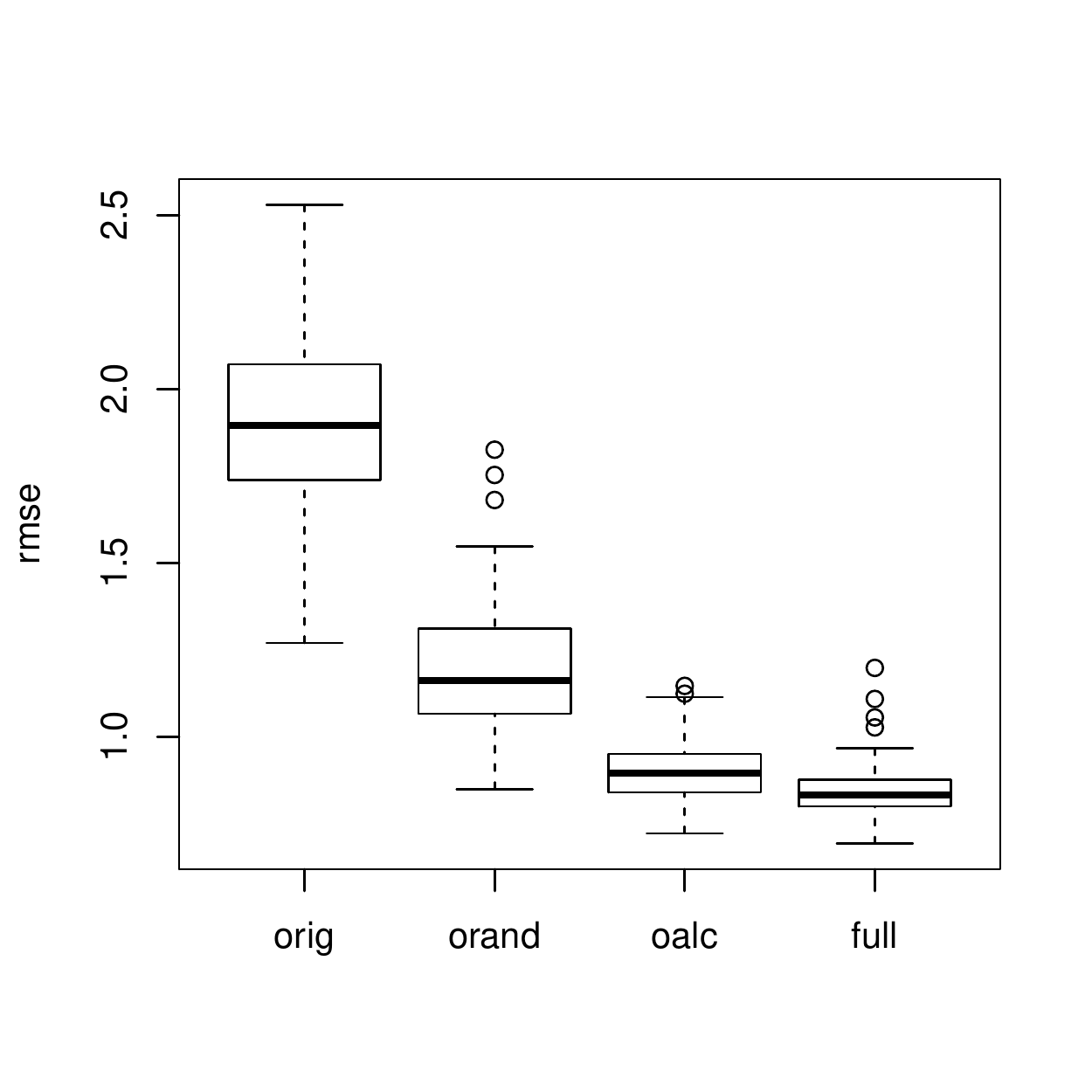}
\end{minipage}
%\hspace{0.5cm}
\begin{minipage}{5cm}
\footnotesize
\begin{tabular}{l|rrr}
& \multicolumn{3}{c}{average predictive density}\\
 &          mean &       5\%   &    95\% \\
\hline
ORIG & 0.15493 & 0.13321 & 0.17607 \\
ORAND & 0.22973 & 0.20596 & 0.25416\\
OALC & 0.25695 & 0.23897 & 0.27492\\
FULL & 0.27116 & 0.25368 & 0.28242\\
% \end{tabular}
% \begin{tabular}{l|rrr}
& \multicolumn{3}{c}{rmse}\\
%    &       mean       & 5\%   &    95\% \\
\hline
ORIG & 1.91487 & 1.55182 & 2.31809 \\
ORAND & 1.19120 & 0.95689 & 1.45869 \\
OALC & 0.89582 & 0.75444 & 1.04228 \\
FULL & 0.84398 & 0.73705 & 0.96311 
\end{tabular}
%\vspace{0.25cm}
\end{minipage}
\caption{Friedman data comparisons by average posterior predictive density
  (higher is better) and RMSE (lower is better).}
\label{f:fried}
\end{figure}

Figure \ref{f:fried} reveals that random retiring is better than subsetting, but retiring by ALC is even better, and can be nearly as good as the full-data estimator.  In fact, OALC was the {\em best} predictor {16\%} and {28\%} of the time by average predictive density and RMSE, respectively.  The average time used by each estimator was approximately 1, 33, 45, and 67 seconds, respectively.  So random retiring on this modestly-sized problem is 2-times faster than using the full data.  ALC costs about 18\% extra, time-wise, but leads to about a 35\% reduction in RMSE relative to the full estimator.  We note that in much larger problems the gap between the online and full estimators can widen considerably.  The time-demands of the full estimator grow roughly as $t\log t$, whereas the online versions stay constant.

\subsubsection*{Spam classification data}

Now consider the Spambase data set, from the UCI Machine Learning Repository \citep{Asuncion+Newman:2007}. The data contains binary classifications of 4601 emails based on 57 attributes (predictors). We report on a similar experiment to the Friedman/regression example, above, except with classification leaves and 5-fold CV to create training and testing sets.  This was repeated twenty times, randomly, giving 100 sets total. Again, four estimators were used: one based on 1/10 of the training fold (ORIG), one based on the full fold (FULL), and two online versions trained on the same stream(s) using either random (ORAND) or entropy (OENT) retiring to keep the total active data set limited to 1/10 of the full set.  

\begin{figure}[ht!]  \centering \begin{minipage}{8.5cm} \includegraphics[scale=0.35,trim=20 50 0 50]{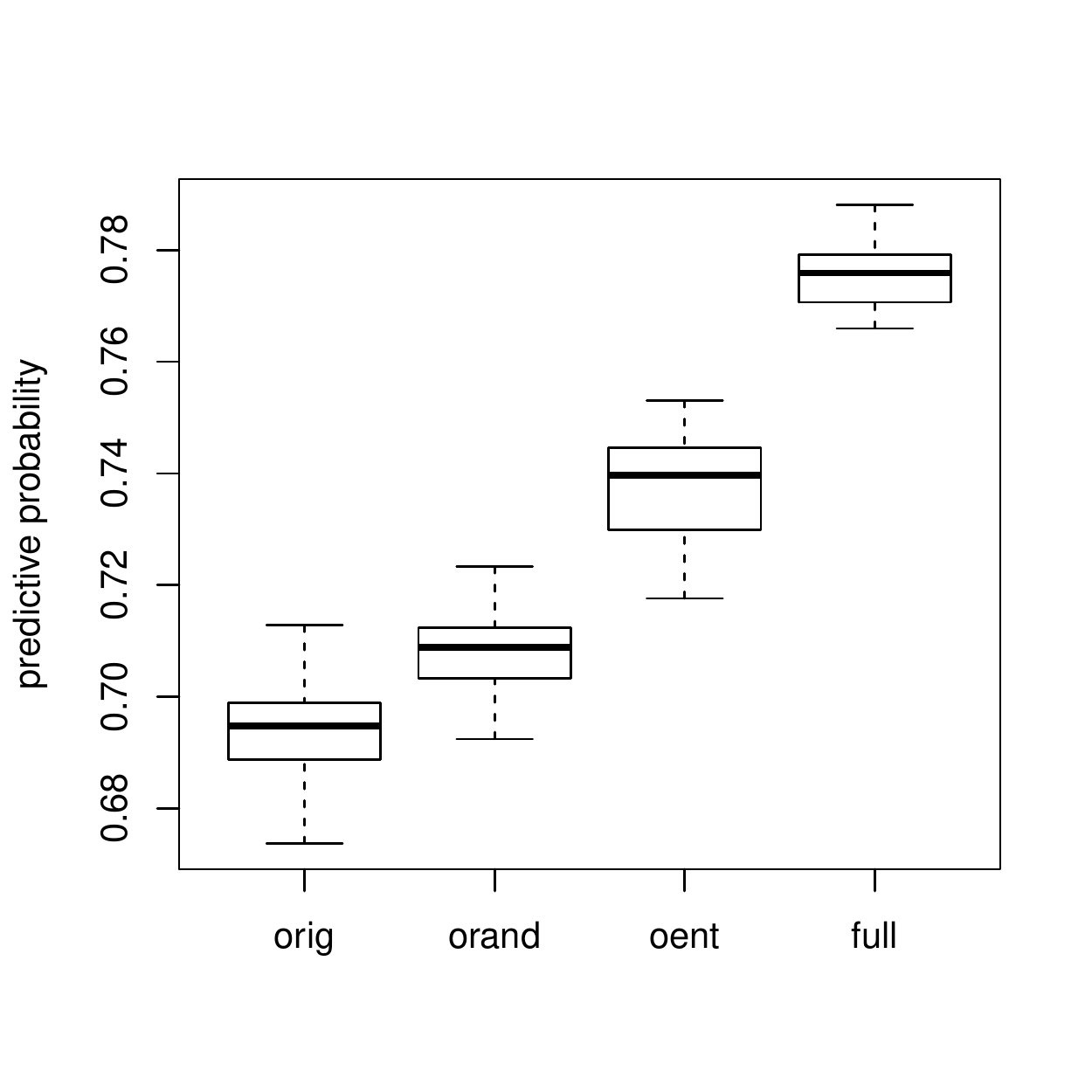} \includegraphics[scale=0.35,trim=20 50 0 50]{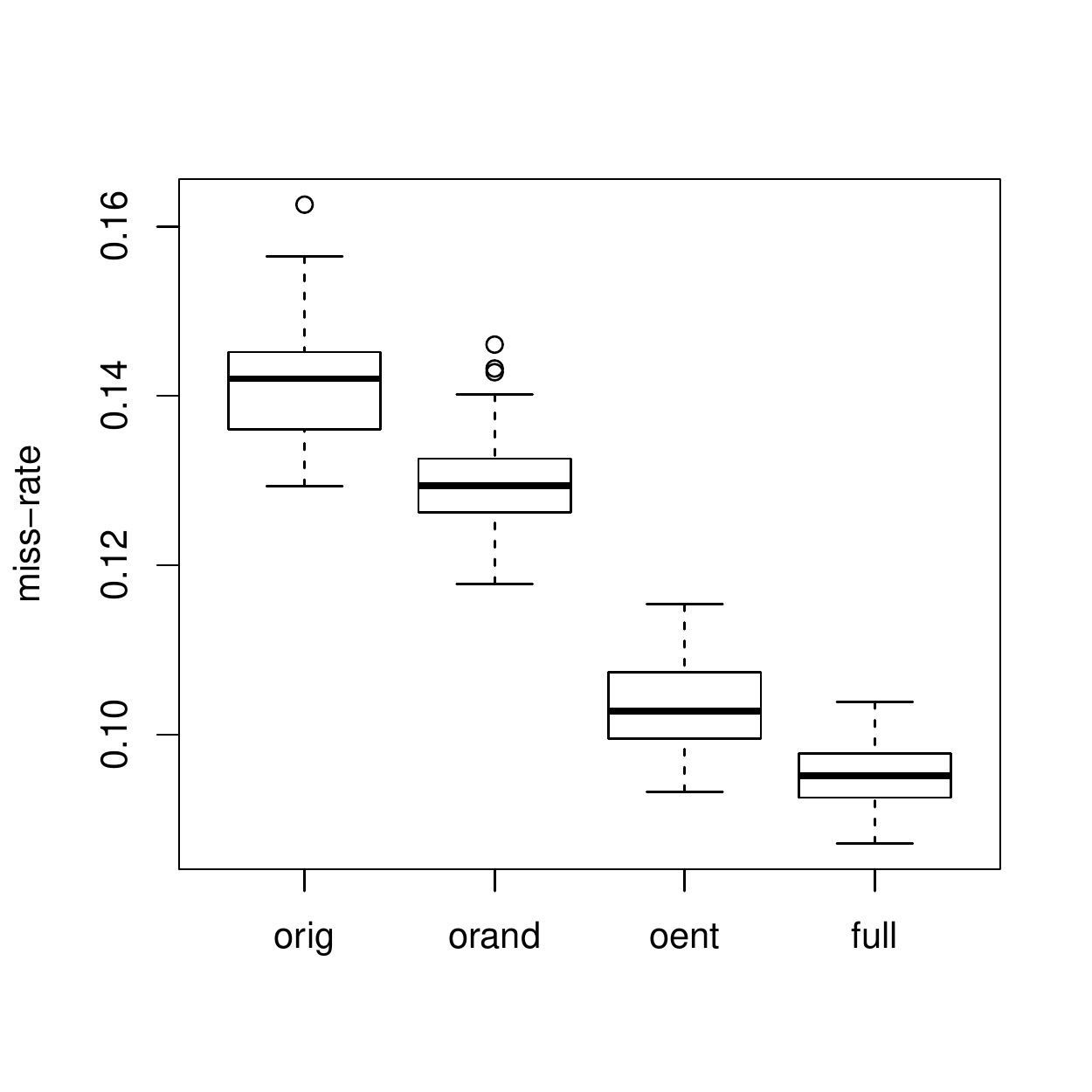} \end{minipage}
%\hspace{0.5cm}
\begin{minipage}{5.5cm}
\footnotesize
%\vspace{-0.25cm}
\begin{tabular}{l|rrr}
& \multicolumn{3}{c}{average predictive probability}\\
    &       mean       & 5\%   &    95\% \\
\hline
ORIG & 0.69393 & 0.68103 & 0.70872 \\
ORAND & 0.70824 & 0.69651 & 0.72135 \\
OENT & 0.73761&  0.72290 & 0.75019\\
FULL  & 0.77620 & 0.76968 & 0.78608\\
%\\
& \multicolumn{3}{c}{misclassification rate}\\
\hline
ORIG     &  0.14109 & 0.13014 & 0.15345 \\
ORAND & 0.13049 &0.12162 & 0.14303 \\
OENT   & 0.10294 &0.09433 & 0.10968 \\
FULL    & 0.09518 & 0.08807 & 0.10195 
\end{tabular}
%\vspace{0.25cm}
\end{minipage}
\caption{Spam data comparisons by average posterior predictive probability
  (higher is better) and
  misclassification rate (lower is better) on the testing set(s).}
\label{f:spam}
\end{figure}

Figure \ref{f:spam} tells a similar story to the Friedman experiment: random discarding is better than subsetting, but discarding by entropy is even better, and can be nearly as good as the full-data estimator.  Entropy retiring resulted the best predictor 7\% of the time by misclassification rate, but never by posterior predictive probability.  % Observe that while ORAND predicts the label about as well as ORIG, it is much better at accurately capturing the full predictive distribution.

\section{Temporal adaptivity using forgetting factors}\label{sec:forget}
The accumulation of historical information at the leaf priors introduced by data retirement may eventually overpower the likelihood of active datapoints. This is natural in an i.i.d. setting, but may cause performance deterioration in streaming contexts where the data generating mechanism may evolve or change suddenly.  To promote responsiveness, we may \emph{exponentially downweight} the retired data history ${s}$ when retiring an additional point $y_m$: $ %\begin{align}
\notag \pi_\lambda^{(\text{new})}(\theta) \propto L(\theta \mid y_m) L^\lambda (\theta; (y_s,\bm{x}_s)_{\{s\}}) \pi_0(\theta).  $ % \end{align}
Observe that for $\lambda = 1$, conjugate Bayesian updating is recovered, and for $\lambda = 0$, the retired history is disregarded altogether, effectively resetting the prior. For $\lambda \in (0,1)$, two effects are introduced. First, the overall `strength' of the prior relative to the likelihood is diminished. Second, as the prior is sequentially updated, it will place disproportionately more weight on recently retired datapoints as opposed to older retired data. For the leaf models entertained in this paper, a recursive application of this principle, with $\lambda \in (0,1)$, modifies only slightly the conjugate updates of Section \ref{sec:retire}, as follows. For the linear and constant models, we have $\left(\bm{A}^{\text{(new)}}\right)^{-1} = \lambda \bm{A}^{-1} + X_m' X_m$, $\bm{R}^{(\text{new})} = \lambda \bm{R} + \X_m'\y_m$, $s^{\text{(new)}} = \lambda s + y_m'y_m$, and $\nu^{\text{(new)}} = \lambda \nu + 1$, whereas for the multinomial, we get $\bm{a}^{(new)} = \lambda \bm{a} + z_m$. For $\lambda < 1$, $\kappa$ and $\nu$ will be bounded above by their limiting value $\frac{1}{1-\lambda}$, irrespective of the total number of retired datapoints. 
%Bounding the strength of informative priors in this way allows model to be responsive to evolving data characteristics.

In \cite{ibrahim2003oop}, this family of priors is shown to satisfy desirable information-theoretic optimality properties. Exponential downweighting as a means of enabling temporal adaptivity also has a long tradition in non-stationary signal processing \citep{haykin1996aft}, as well as streaming classification \citep{anagnostopoulos2012}, where $\lambda$ is often referred to as a \emph{forgetting factor}.

In historical discarding, it is perhaps obvious that some degree of forgetting will be useful in drifting contexts, as the contribution of past data becomes decreasingly useful with time. The relationship between forgetting and other types of active discarding is however more complex. In principle, any successful active discarding scheme will lead to priors being populated by less relevant datapoints, so that the model can benefit from forgetting in favour of putting more weight on highly relevant, active data. Unfortunately, in the presence of drift, we cannot guarantee such reasonable behaviour from active discarding heuristics of the form proposed here. As these latter are reliant on an i.i.d. assumption, they can often mistake obsolete datapoints that are poorly explained by the model for `highly informative, surprising' datapoints that had better be retained, so that it becomes less clear \emph{a priori} whether the active data pool or the prior should be `trusted' more, and the utility of forgetting becomes questionable. More sophisticated active learning heuristics are required to resolve this problem, which lie beyond the scope of this paper. We will thus only explore the interaction of forgetting with historical discarding henceforth. 

\subsection{Synthetic drifting regression data}
We now revisit the Friedman dataset from \ref{sec:adresults}, and introduce smooth drift by replacing the non-linear term $10 \sin(\pi x_1 x_2) $ with a time-varying version, $10 a_t \sin(\pi x_1 x_2)$. The coefficient $a_t$ is allowed to vary smoothly between $-1$ and $3$ over time as $a_t = 2 \sin(2 \pi k t/1000)+1$, so that $k$ controls the speed of the drift: $k=1$ producing one full cycle every 1000 timesteps. Note that as $a_t$ increases in magnitude, the non-linearity of the regression surface will accentuate, as the first term is responsible for much of its complexity. The simulation measures $1$-step-ahead performance of the DT as follows: at each timestep $t$, it first generates $5$ datapoints from the current model; these are used as test datapoints to measure the predictive probability and RMSE of the dynamic tree (trained using data up to time $t-1$); and, finally, the DT is updated on the basis of the new data.

%Consecutive snapshots of $y$ against $X_1$ in the top plot of Figure \ref{fig:results_FRIED_H} clearly demonstrate this effect. 
\begin{figure}[ht!]
\begin{minipage}{\textwidth}
\centering
\includegraphics[width=0.45\textwidth,trim=30 40 30 10, clip=true]{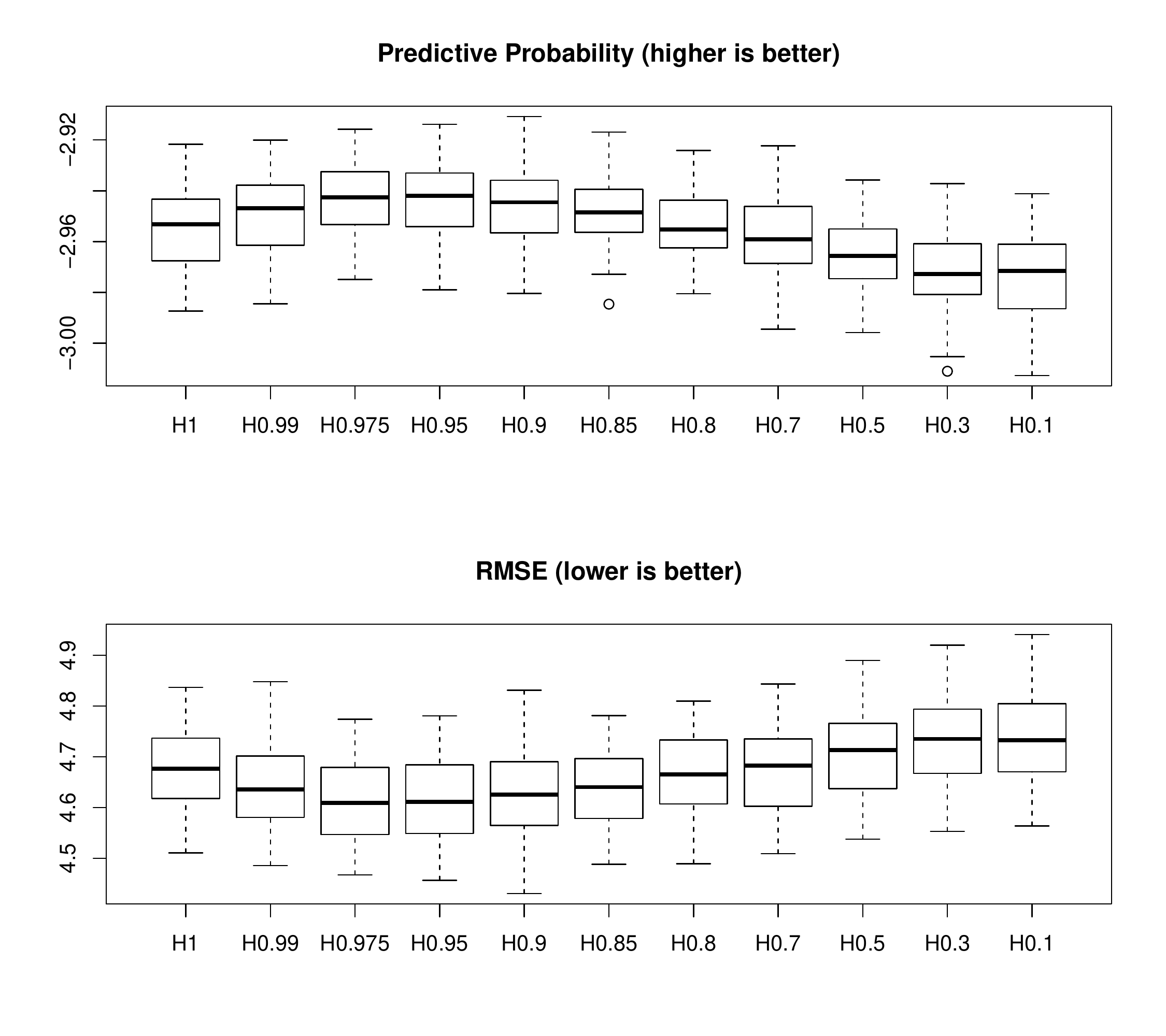}
\includegraphics[width=0.45\textwidth,trim=30 40 30 20, clip=true]{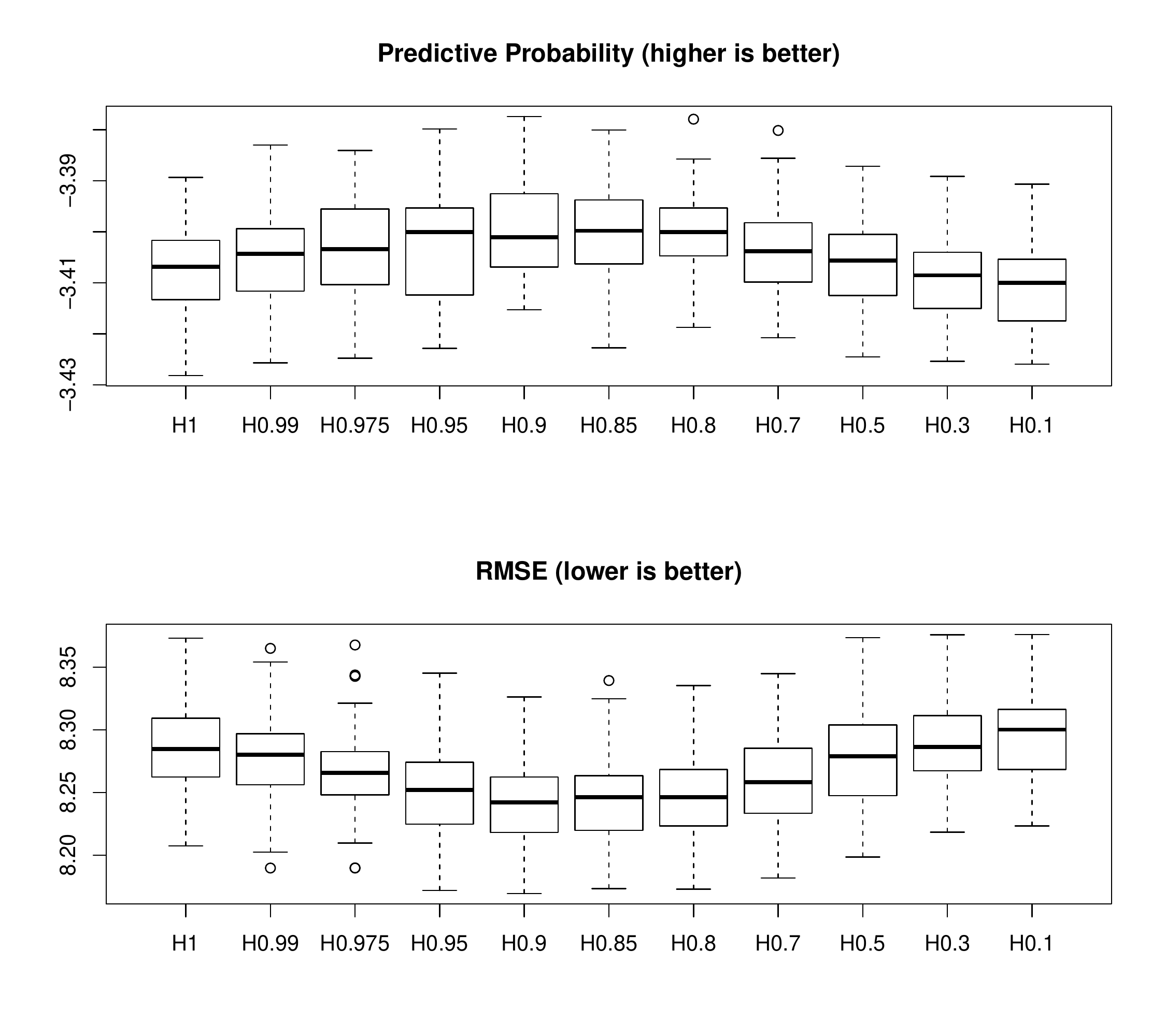}
\end{minipage}
\caption{%Top: snapshots of the regression relationship between $y$ and the product of the first covariate, $x_1$, for six consecutive time intervals. 
Slowly (left) and rapidly (right) drifting Friedman data comparisons by average posterior predictive density (top - higher is better) and RMSE (bottom - lower is better), for various degrees of forgetting.}
\label{fig:results_FRIED_H}
\end{figure}

In the Figure \ref{fig:results_FRIED_H} plots, we plot the RMSE and predictive probability observed over 100 MC iterations for a sequence of $\lambda$ values between $\lambda = 0$ (discarding with retiring) and $\lambda = 1$ (retirement via Bayesian conjugate updating).
% against data simulated from the drifting Friedman model described above, for a DT with historical discarding.
%Recall that $\lambda=1$ corresponds to standard Bayesian conjugate updating, and $\lambda=0$ corresponds to discarding without retiring. % (i.e., datapoints removed from the active memory are discarded altogether, and not entered into the prior).
Reassuringly, a U-shaped curve appears, indicating a trade-off between throwing away too much information at one extreme ($\lambda = 0$), and retaining obsolete information at the other ($\lambda = 1$). For rapidly changing data distributions ($k=1$), a value of $\lambda = 0.8$ seems to perform best. Repeating the experiment for slower-changing data distributions ($k = 0.1$) produces performance that peaks at $\lambda = 0.97$ instead, confirming our intuition.

\begin{figure}[ht!]
 \centering
\includegraphics[width=0.45\textwidth, trim= 0 0 0 0]{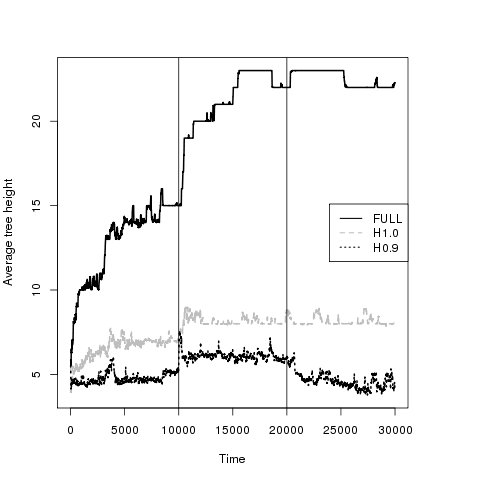}
\caption{Average tree height over time for the full model, and historical discarding with $\lambda=0.9$ or $\lambda=1$.  The true regression surface complexity rises in $t\in [10^4,2\times 10^4]$.}
\label{fig:heightpath}
\end{figure}

In Figure \ref{fig:heightpath}, we investigate the effect that discarding and forgetting have on model complexity, as measured by average tree height over time. To do so, we again generate drifting Friedman data, this time with $a_t=10$ between $t=10000$ and $t=20000$ (denoted by vertical lines in Figure \ref{fig:heightpath}), and $0$ otherwise so that model complexity rises sharply and then drops again. We deploy a DT without discarding (i.e., sequentially incorporating the full dataset), a DT with a fixed budget of $100$ active datapoints and no forgetting, and one with the same budget and mild forgetting ($\lambda = 0.9$).  First observe that capping the active data pool size significantly penalises model complexity on the whole. Also note that all three methods react to the rise in complexity at $t=10^4$ by favouring deeper trees. However, once the data complexity drops again at $t=2\times 10^4$, both the full model and the online model without forgetting ($\lambda=1$) retain their average tree depth, failing to return to earlier levels.  By contrast, $\lambda=0.9$ allows the model to adapt to the change, as the priors are more easily outweighed by the impact of novel information.

\subsection{Synthetic drifting classification data}
We now turn to streaming classification. We henceforth adopt the standard one-step-ahed performance assessment paradigm, wherein the algorithm at time $t$ first predicts the class label of the unlabelled $(t+1)$th datapoint, and is then allowed to use both the datapoint itself and its label to update its parameters.

\begin{figure}[ht!]
\begin{minipage}{\textwidth}
\centering
\includegraphics[width=0.4\textwidth]{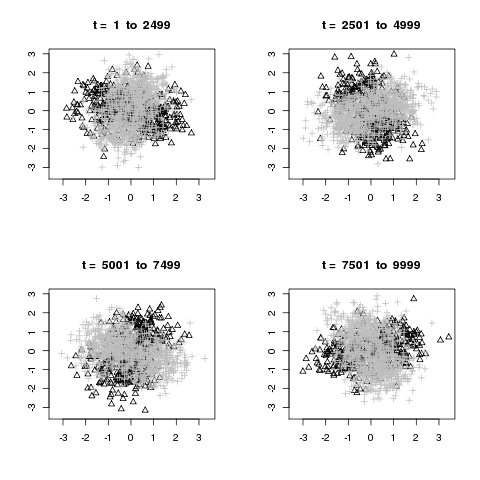} \ \ \ 
\includegraphics[width=0.43\textwidth,trim=0 0 0 20]{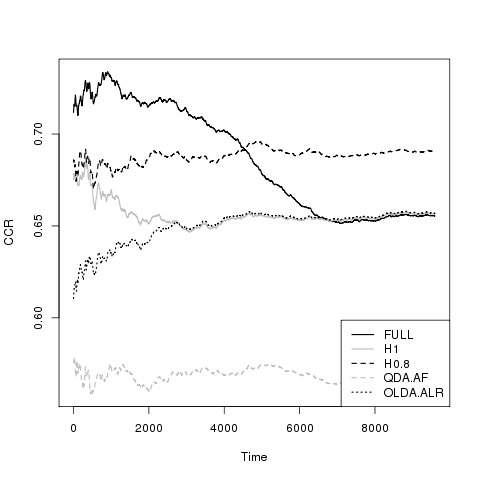}
\end{minipage}
\caption{Left: four snapshots of the drifting synthetic classification data underlying this simulation study. Right: average classification performance (Correct Classification Rate) over time.}
\label{fig:snap_class}
\end{figure}

We first consider a classification problem where the optimal decision boundary is always non-linear, but drifts in time in such a way so that older data become increasingly misleading for future predictions. This effect can be synthesised by rotating a `fuzzy' XOR problem, displayed in the left plot of Figure \ref{fig:snap_class}. The XOR forces a non-linear decision boundary, whereas the rotation implies that recent data should have priority over older data. This example, which we refer to as MOVINGTARGET, is an extreme one since in general drift could also manifest itself in ways that render past information useless, but not outright misleading. Even in such cases, data discarding and forgetting may be useful to `free up' degrees of freedom, but the effect is unlikely to be as dramatic and would therefore be harder to measure.

In the right plot of Figure \ref{fig:snap_class} and the rightmost column of Table \ref{table:results}, we illustrate the effect that introducing a forgetting factor has on the performance of a DT with historical discarding against MOVINGTARGET. We compare against two state-of-the-art methods, Quadratic Discriminant Analysis with Adaptive Forgetting (QDA-AF) \citep{anagnostopoulos2012}, and Online Linear Discriminant Analysis with Adaptive Learning Rate (OLDA-ALR) \citep{kuncheva2008alr}, both designed with streaming classification contexts in mind.  In Table \ref{table:results}, performance is measured in three ways: correct classification rate, Area Under the Curve, the $H$-measure, a newly preferred alternative to the AUC \citep{hand2009}. In all three respects, data discarding hardly improves performance when $\lambda=1$, whereas for $\lambda=0.9$ significant improvement is possible. For an explanation, consider the way in which classification performance evolves over time, shown on the rightmost plot of Figure \ref{fig:snap_class}: the detrimental effect of obsolete, misleading data becomes visually obvious for the full-data model, as well as the online model with $\lambda=1$. Interestingly, these latter two methods, although otherwise distinct, share a similar performance bottleneck with OLDA-ALR that DTs with forgetting to decidedly overcome.

\begin{table}
\caption{Performance in terms of Area Under the Curve (AUC), H-measure (H) and Correct Classification Rate (CCR) for each of three datasets (two real and one simulated), and three instantiations of dynamic trees as well as two competitive methods. \label{table:results}}
\begin{tabular}{l|rrr|rrr|rrr}
\bf & \multicolumn{3}{c}{ELEC2} & \multicolumn{3}{|c}{FAUD} & \multicolumn{3}{|c}{ MOVINGTARGET }\\
\bf &  AUC & H & CCR & AUC & H & CCR & AUC & H & CCR\\\hline
OFFLINE ($n=10^4$) 	&  0.771 & 0.267 & 0.702 & 0.588 & 0.048 & 0.941 & 0.562 & 0.044 & 0.655\\
ONLINE ($\lambda = 1$) 	&  0.761 & 0.274 & 0.724 & 0.724 & 0.155 & 0.971 & 0.528 & 0.011 & 0.656\\
ONLINE ($\lambda = 0.8$)&  0.880 & 0.480 & 0.808 & 0.930 & 0.622 & 0.982 & 0.668 & 0.111 & 0.609\\
QDA.AF 			&  0.924 & 0.643 & 0.873& 0.973 & 0.920 & 0.983 & 0.504 & 0.001 & 0.560\\
OLDA.ALR 		&  0.763 & 0.239 & 0.692& 0.832 & 0.414 & 0.974 & 0.507 & 0.001 & 0.656
\end{tabular}
\end{table}

\begin{figure}[ht!]
\centering
\includegraphics[width=0.45\textwidth,trim=0 0 0 25]{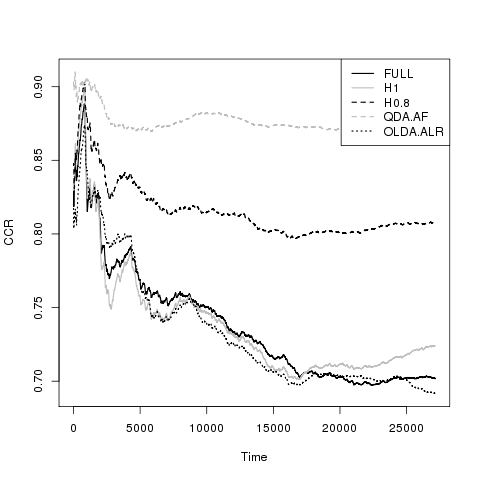}
\includegraphics[width=0.45\textwidth,trim=0 0 0 25]{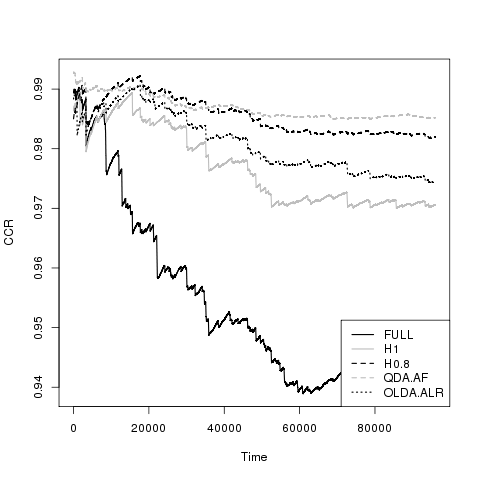}
\caption{Left: Electricity Market data. Right: Fraud data. Average classification performance (Correct Classification Rate) over time.}
\label{fig:ELEC2}
\end{figure}

Now consider two real datasets, ELEC2, and FRAUD, which are known to exhibit concept drift \citep{anagnostopoulos2012}. The former holds information for the Australian New South Wales Electricity Market and was introduced in \cite{baenagarcia2006edd}, comprising $27552$ instances, each referring to a period of $30$ minutes. The class label identifies the price change related to a moving average of the last 24 hours, and the four covariates capture aspects of electricity demand and supply. The latter dataset, FRAUD, is of length prohibitive to many existing methods ($n=100,000$), and contains information about credit card transactions, and their respective status as legitimate or fraudulent, determined by experts (see \cite{anagnostopoulos2012} for more details). % Both are temporal datasets, known to exhibit temporal variation of unknown characteristics. 
Results in terms of CCR over time are presented in Figure \ref{fig:ELEC2}. 
The full suite of numerical results is provided in Table \ref{table:results}.  QDA-AF, which was vastly inferior in the synthetic MOVINGTARGET example, dominates in these examples (particularly for FRAUD). Comparisons between the remaining methods paint an interesting picture. In both datasets, FULL is among the lowest performers, suggesting that data discarding is essential to maintaining a representative fit against drifting data distributions. Among DTs, $\lambda=1$ performs poorly, as informative priors accumulate irrelevant or misleading information. In contrast, $\lambda=0.8$ is much better, outperforming OLDA-ALR. % and is competitively with the best method, especially for the FRAUD data.
Although the ranking of various classifiers will generally differ by application, these experiments give a strong signal that the use of forgetting factors can turn DTs into a promising, flexible tool for streaming classification.

\section{Conclusion}
\label{sec:conclude}
In this work, we strive to fully utilise the potential of Bayesian machinery in the context of {streaming} non-parametrics. We propose data retirement via conjugate Bayesian updating in the context of SMC inference for a dynamic tree model, preserving non-parametric flexibility while enabling constant memory online operation. Second, the availability of tractable predictive distributions allows us to devise computationally efficient active retirement heuristics, hence maintaining a fixed budget of highly informative datapoints. Both features minimise information loss incurred by single-pass processing.  {Finally, we deploy informative power priors to enable temporal adaptivity.} This results in a novel, powerful algorithmic scheme for non-parametric regression and classification tailored to the massive and streaming data contexts.  As future work, we intend to pursue techniques for automatic tuning of forgetting factors in streaming contexts, and their interplay with active retirement heuristics.

\section{Acknowledgements}
The first author was supported by a Cambridge Statistics Initiative Research Fellowship at the Statistical Laboratory, University of Cambridge, for a large part of this work. 

\bibliographystyle{Chicago}

\bibliography{trees}

\end{document}